\documentclass{amsart}
\date{March 2, 2000}
\DeclareMathOperator{\sgn}{sgn} 
\DeclareMathOperator{\arsinh}{arsinh} 

\newcommand{\gH}{\mathfrak{H}}

\newcommand{\bg}{\mathbf{g}}
\newcommand{\bp}{\mathbf{p}}
\newcommand{\bq}{\mathbf{q}}
\newcommand{\bx}{\mathbf{x}}
\newcommand{\by}{\mathbf{y}}

\newcommand{\cz}{\mathbb{C}} 
\newcommand{\rz}{\mathbb{R}} 

\newcommand{\bH}{\mathbb{H}^{\mathrm{bare}}}

\newtheorem{lemma}{Lemma}

\newtheorem{theorem}{Theorem}
\newtheorem{corollary}{Corollary}

\title[QED]{Renormalization of the Regularized Relativistic Electron-Positron
  Field}

\author[E.H.~Lieb]{Elliott H.~Lieb}
\address{Departments of Mathematics and Physics\\
  Jadwin Hall\\ 
  Princeton University\\
  P.O.B. 708\\
  Princeton, NJ 08544-0708\\ USA} \email{lieb@princeton.edu}
\author[H.~Siedentop]{Heinz Siedentop}
\address{Mathematik I\\
  Universit\"at Regensburg\\
  93040 Regensburg\\
  Germany} \email{heinz.siedentop@mathematik.uni-regensburg.de}

\begin{document}
\begin{abstract} 
We consider the relativistic electron-positron field interacting with
itself via the Coulomb potential defined with the physically motivated,
positive, density-density quartic interaction. The more usual
normal-ordered Hamiltonian differs from the bare Hamiltonian by a
quadratic term and, by choosing the normal ordering in a suitable,
self-consistent manner, the quadratic term can be seen to be equivalent
to a renormalization of the Dirac operator. Formally, this amounts to
a  Bogolubov-Valatin transformation, but in reality it is
non-perturbative, for it leads to an inequivalent, fine-structure
dependent representation of the canonical anticommutation relations.
This non-perturbative redefinition of the electron/positron states can
be interpreted as a mass, wave-function and charge renormalization,
among other possibilities, but the main point is that a
non-perturbative definition of normal ordering might be a useful
starting point for developing a consistent quantum electrodynamics.

\end{abstract}

\footnotetext[1]{
\copyright\,1999 by the authors. This paper may be reproduced, in its
entirety, for non-commercial purposes.}


\maketitle

\section{Introduction\label{S1}}

In relativistic quantum electrodynamics (QED) the
quantized electron-positron field $\Psi(x)$, which is an operator-valued
spinor, is written formally as 
\begin{equation}\label{psi}
\Psi(x):= a(x) + b^*(x), 
\end{equation}
where $a(x)$ annihilates an electron at $x$ and $b^*(x)$ creates a
positron at $x$.  (We use the notation that $x$ denotes a space-spin
point, namely $x = (\mathbf x, \sigma)\in \Gamma=\rz^3\times
\{1,2,3,4\}$ and $\int d^3x$ denotes integration over $\rz^3 $ and a
summation over the spin index.)  More precisely, we take the Hilbert
space, $\gH$ of $L^2(\rz^3)$ spinors, i.e., $\gH=L^2(\rz^3)\otimes
\cz^4$. To specify the one-electron space we choose a subspace,
$\gH_+$ of $\gH$ and  denote the orthogonal projector onto this
subspace by $P_+$. The one-positron space is the (anti-unitary) charge
conjugate, $C\gH_-$, of the orthogonal complement, $\gH_-$ with
projector $P_-$.  In position space, $(C(\psi))(\bx):=
i\beta\alpha_2\overline{\psi(\bx)}$, whereas in momentum space
$(C(\hat\psi))(\bp):= i\beta\alpha_2\overline{\hat\psi(-\bp)}$.  For
later purposes we note explicitly that the Hilbert space $\gH$ can be
written as the orthogonal sum
\begin{equation}
  \label{eq:orth}
  \gH=\gH_+\oplus\gH_-.
\end{equation}
If $f_\nu$ is an orthonormal basis for $\gH_+$ with $\nu \geq 0$ and
an orthonormal basis for $\gH_-$ with $\nu < 0$ then $a(x):=\sum_{\nu
  =0}^\infty a(f_\nu) f_\nu(x)$, and $b(x):=\sum_{\nu=-\infty}^{-1}
b(f_\nu) \overline{f_\nu(x)}$, where $a^*(f)$ creates an electron in
the state $P_+ f$, etc.  (For further details about the notation see
Thaller \cite{Thaller1992}, and also Helffer and Siedentop
\cite{HelfferSiedentop1998} and Bach et al
\cite{Bachetal1999,Bachetal1998}.

For free electrons and positrons, $\gH_+$ and $\gH_-$ are,
respectively, the positive and negative energy solutions of the free
Dirac operator
\begin{equation}\label{dirac}
  D_0 = \boldsymbol{\alpha} \cdot \mathbf{p} +m_0 \beta ,
\end{equation}
in which $\boldsymbol{\alpha}$ and $\beta$ denote the four $4\times 4$
Dirac matrices and $\bp=-i \nabla$.   The
number $m_0$ is the \textit{bare mass} of the electron/positron.
Perturbation theory is defined in terms of this splitting of $\gH$
into $\gH_+$ and $\gH_-$.  The electron/positron lines in Feynman
diagrams are the resolvents of $D_0$ split in this way. As we shall
see, this splitting may not be the best choice, ultimately, and
another choice (which will require a fine-structure dependent,
inequivalent representation of the canonical anticommutation
relations) might be more useful.

The Hamiltonian for the free electron-positron field is
$$
\mathbb{H}_0= \int d^3x  :\Psi(x)^* D_0 \Psi(x):
$$
with this choice of $\gH_+$. The symbol $:\ :$ denotes normal
ordering, i.e., anti-commuting all $ a^*, b^*$ operators to the left
(but ignoring the anti-commutators).

In this paper we investigate the effect of the Coulomb interaction
among the particles, which is a quartic form in the $a^\#, b^\#$
operators. ($\#$ denotes either a star or no star.) The normal
ordering will give rise to extra constants and certain quadratic
terms.  Our main point will be that by making an appropriate choice of
$\gH_+$ that is different from the usual one mentioned above, we can
absorb the additional quadratic terms into a mass, wave function, and
charge renormalization. This choice of $\gH_+$ has to be made self
consistently and, as we show, can be successfully carried out if some
combination of the fine structure constant $\alpha $ (=1/137 in
nature) and the ultraviolet cutoff $\Lambda$ is small enough (but
`small' is actually `huge' since the condition is, roughly, $\alpha
\log \Lambda <1$).

The physical significance of our construction is not immediate, but it
does show that certain annoying terms, when treated
\textit{non-perturbatively}, can be incorporated into the
renormalization program. It is not at all clear that what we have
called wave function renormalization, for example, really corresponds
to a correct interpretation of that renormalization, or even what the
true meaning of wave function renormalization is. It could also be
interpreted as a renormalization of Planck's constant. Likewise, the
charge renormalization we find is not mandatory. But our main point is
that the effect is mathematically real and is best dealt with by a
change of the meaning of the one-particle electron/positron states. It
should be taken into account in the non-perturbative QED that is yet
to be born.  Of course, there are other renormalization effects in QED
that we do not consider. In particular, the magnetic field has not
been included and so we do not have a Ward identity to help us.  Our
result here is only a small part of the bigger picture of a proper,
non-perturbative QED.

The last section of this paper contains a brief discussion of some
possible interpretations of our findings and the interested reader is
urged to look at that section.

Our starting point is the unrenormalized (`bare') Hamiltonian for a
quantized electron/positron field interacting with itself via
Coulomb's law, namely
\begin{multline}
  \label{eq:1} \mathbb{H}^{\rm bare}\\ 
  := \int d^3x :\Psi^*(x)D_{0}\Psi(x): + \frac\alpha2 \int d^3x \int
  d^3y W(\bx,\by) :\Psi^*(x) \Psi(x)::\Psi^*(y) \Psi(y): \ .
\end{multline}
where $W$ is a symmetric ($W(x,y)=W(y,x)$) interaction potential. The
case of interest here is the Coulomb potential
$W(x,y)=\delta_{\sigma,\tau}|\bx-\by|^{-1}$ and regularized versions
thereof. We will refer to the first term on the right side of
\eqref{eq:1} as the kinetic energy $\mathbb{T}$ and to the second term
as the interaction energy $\mathbb{W}^{\rm bare}$.

Our choice of the product of the two normal ordered factors in the
interaction, namely $:\rho(x):$ $:\rho(y):$, is taken from the book of
Bjorken and Drell \cite{BjorkenDrell1965} (Eq. (15.28) without
magnetic field). It is quite natural, we agree, to start with the
closest possible analog to the classical energy of a field. Of course,
this term is only partly normal ordered in the usual sense and
therefore it does not have zero expectation value in the unperturbed
vacuum or one-particle states.  It is positive, however, as a Coulomb
potential ought to be.

The fully normal ordered interaction (which is not positive) is
defined to be
\begin{equation} \label{normal}
  \mathbb{W}^{\mathrm{ren}}_\alpha 
  =  \frac\alpha2\int d^3x \int d^3y W(\bx,\by)
  :\Psi^*(x) \Psi^*(y) \Psi(y) \Psi(x):  .  
\end{equation} 
Bear in mind that this definition entails a determination of the
splitting \eqref{eq:orth} of $\gH$.

We introduce a normal ordered (`dressed') Hamiltonian
\begin{equation}
  \label{ordered} \mathbb{H}_{\alpha}^{\rm ren} := \int d^3x
  :\Psi^*(x)D_{Z,m} \Psi(x):  +\mathbb{W}^{\mathrm{ren}_\alpha} .
\end{equation} 
(Again, this depends on the splitting of $\gH$.) The operator
$D_{Z,m}$ differs from $D_0$ in two respects:
\begin{equation}\label{newdirac} D_{Z,m}= Z^{-1} (\boldsymbol{\alpha}
  \cdot \mathbf{p} +m \beta ), 
\end{equation} 
where $Z<1 $ and $m > m_0$. We call $m_0$ the bare mass and $m$ the
`physical' mass --- up to further renormalizations, which we do not
address in this paper.  The factor $Z$ is called the `wave function
renormalization' because the `wave function renormalization' in the
standard theory is defined by the condition that the electron-positron
propagator equals $Z$ times a non-interacting propagator, which is the
inverse of a Dirac operator. To the extent that
$\mathbb{W}^{\mathrm{ren}}_\alpha$ can be neglected, the propagator
with just the first term on the right side of \eqref{ordered} would,
indeed, be $Z$ times a Dirac operator with mass $m$. One could also
interpret $Z$ in other ways, e.g., as a renormalization of the speed
of light, but such a choice would be considerably more radical. Some
other interpretations are discussed in the last section.

Our goal is to show that by a suitable choice of $Z$, $m$, and the
electron space $\gH_+$, we have -- up to physically unimportant
infinite additive constants -- asymptotic equality between $
\mathbb{H}^{\mathrm{bare}}$ and $\mathbb{H}^{\mathrm{ren}}_\alpha$ for
momenta that are much smaller than the cutoff $\Lambda$.
\begin{equation}
\mathbb{H}^{\mathrm{bare}} =  \mathbb{H}^{\mathrm{ren}}_\alpha
\quad\quad \quad \mbox{ (for small momenta)}
\end{equation}

The additional normal ordering in \eqref{ordered} will introduce some
quadratic terms and it is these terms to which we turn our attention
and which we would like to identify as renormalization terms. Whether
our picture is really significant physically, or whether it is just a
mathematical convenience remains to be seen. What it does do is
indicate a need for starting with a non-conventional, non-perturbative
choice of the free particle states, namely the choice of $\gH_+$.

\section{Calculation of the New Quadratic Terms \label{calc}}
The canonical anti-commutation relations are
\begin{equation}
  \begin{split}
    \{a(f),a(g)\}  &= \{a^*(f),a^*(g)\}=\{a(f),b(g)\}=\{a^*(f),b^*(g)\}\\
                   &=\{a^*(f),b(g)\}=\{a(f),b^*(g)\}=0, 
  \end{split}
  \label{car1} 
\end{equation}
and
\begin{equation}
  \{a(f),a^*(g)\}= (f,P_{+}g),\ \{b^*(f),b(g)\}= (f,P_{-}g).
  \label{car2} 
\end{equation} 
Formally, this is equivalent to
\begin{equation}
  \begin{split}
    \{a(x),a(y)\}  &= \{a^*(x),a^*(y)\}=\{a(x),b(y)\}=\{a^*(x),b^*(y)\}\\
    & =\{a^*(x),b(y)\}=\{a(x),b^*(y)\}=0, 
  \end{split}
  \label{car3} 
\end{equation}
and
\begin{equation}
  \{a(x),a^*(y)\}= P_{+}(x,y),\ \{b^*(x),b(y)\}= P_{-}(x,y),
  \label{car4} 
\end{equation} 
where $P_{\pm}(x,y)$ is the integral kernel of the projector
$P_{\pm}$.

In order to compare the bare Hamiltonian $\mathbb{H}^{\mathrm {bare}}$
with the renormalized one $\mathbb{H}^{\mathrm {ren}}_\alpha $ we must
face a small problem; the momentum cutoff
$\Lambda$, which is necessary in order to get finite results, will
spoil the theory at momenta comparable to this cutoff.  Thus, our
identification of the difference as a wave function and mass
renormalization will be exact only for momenta that are small compared
to $\Lambda$. We will also require (although it is not clear if this is
truly necessary or whether it is an artifact of our method) that
$\alpha \ln (\Lambda/ m_0)$ is not too large.

The potential energy $\mathbb{W}^{\rm bare}$ clearly has positive
singularities in it if the unregularized Coulomb interaction
$W(x,y)=|\bx-\by|^{-1}\delta_{\sigma,\tau}$ is taken. We therefore cut
off the field operators for high momenta above $\Lambda$. (Another
choice would be to cut off the Fourier transform of $|\bx-\by|^{-1}$
at $\Lambda$ and the result would be essentially the same; our choice
is motivated partly by computational convenience.) Our cutoff
procedure is to require that
\begin{equation}\label{cutoff}
  \Psi(x) := (2\pi)^{-3/2}\int_{\{\bp\in\rz^3\ |\ |\bp|<\Lambda\}}
  \widehat{\Psi}(\bp,\sigma)e^{-i\bp\cdot\bx} d\bp
\end{equation} 
holds, with $\Lambda >0$. We could also take a smooth cutoff without
any significant change.  This regularization keeps the positivity of
$\mathbb{W}^\mathrm{bare}$. To make all terms finite we also need a
volume cutoff for one of the difference terms. However, the volume
singularity only occurs as an additive constant, which we drop.  The
renormalized Hamiltonian is free of infrared divergent energies and so
does not need a volume cutoff.

Next, we calculate the difference
  \begin{multline}\label{eq:6}
    \mathbb{R}:=
    \mathbb{W}^\mathrm{bare}-\mathbb{W}^\mathrm{ren}_\alpha =
\frac{\alpha}{2} \int
    d^3x \int d^3y\ W(\bx,\by) [a^*(x)a(y) P_+(x,y) +
    a^*(x)b^*(y) P_+(x,y)\\
    +b(x)a(y)P_+(x,y) - b^*(y)b(x)P_+(x,y) \\
    -a^*(y)a(x)P_-(x,y) + P_-(x,y)P_+(x,y)+a(x)b(y)P_-(x,y)\\ +
    b^*(x)a^*(y)P_-(x,y)+b(x)^*b(y)P_-(x,y)] \ .
\end{multline} 
Thus, \eqref{eq:1} can be rewritten as 
\begin{multline}
  \label{eq:7} 
      \bH = \int d^3x :\Psi^*(x)D_0\Psi(x): \\ 
    +\frac{\alpha}{2} \int d^3x
    \int d^3y \frac{P_+(x,y)-P_-(x,y)}{ |\bx-\by|}
    :\Psi^*(x)\Psi(y):  +\mathbb{W}^{\mathrm{ren}}_\alpha\\
    +\frac\alpha2\int d^3x \int d^3y \frac{P_+(x,y)P_-(x,y)}{|\bx-\by|}.
  \end{multline}

The first two terms on the right side are one-particle operators
(i.e., quadratic in the field operators). Let us call their sum
$\mathbb{A}$, which we write (formally, since $A$ is a differential
operator) as
\begin{equation}\label{A} 
  \mathbb{A} = \int d^3x \int d^3y A(x,y):\Psi^*(x)\Psi(y):.  
\end{equation} 
The last term is a cutoff dependent constant, which happens to be
infinite.  (One can show that this term is positive and, if we put in
a momentum cutoff $\Lambda$ and volume (infrared) cutoff $V$, this
integral is proportional to $V\Lambda^4$.)

We can make $\mathbb{A}$ positive by choosing the normal ordering
appropriately. That is, we choose $P_+$ to be the projector onto the
positive spectral subspace of the operator ${A}$. Thus, we are led to
the nonlinear equation for the unknown ${A}$ 
\begin{equation}
  \label{eq:8} 
  A = D_0 + \frac\alpha2 R ,
\end{equation} 
where the operator $R$ has the integral kernel 
\begin{equation} \label{eq:9}
R(x,y) := \frac{P_+(x,y)-P_-(x,y)}{ |\bx-\by|}= \frac{(\sgn A)
(x,y)}{|\bx-\by|}, \end{equation} with $\sgn t $ being the sign of
$t$.

For physical reasons and to simplify matters we restrict the search
for a solution to \eqref{eq:8} to translationally invariant operators,
i.e., $4\times 4$ matrix-valued Fourier multipliers.  Moreover, we
make the ansatz
\begin{equation}
  \label{eq:10} A(\bp):= \boldsymbol\alpha\cdot\omega_\bp g_1(|\bp|)+
  \beta g_0(|\bp|) 
\end{equation} 
with real functions $g_1$ and $g_0$, where $\omega_\bp$ is the unit
vector in the $\bp$ direction.  In other words, we try to make $A$
look as much as possible like a Dirac operator.  With this ansatz we
have $\sgn A(\bp) = A(\bp)/(g_1(|\bp|)^2 + g_0(|\bp|)^2)^{1/2}$. Thus,
recalling \eqref{cutoff}, \eqref{eq:8} can be fulfilled, if
\begin{align}
  \label{eq:12} g_0(|\bp|) &:= m_0 + \frac\alpha{4\pi^2}
  \int_{|\bq|<\Lambda} d\bq \frac{1}{|\bp-\bq|^2}
\frac{g_0(|\bq|)}{ (g_1(|\bq|)^2 + g_0(|\bq|)^2)^{1/2}}\\
  \label{eq:11} g_1(|\bp|) &:= |\bp| + \frac{\alpha}{4\pi^2}
  \int_{|\bq|<\Lambda} d\bq
  \frac{\omega_\bp\cdot\omega_\bq}{|\bp-\bq|^2} 
\frac{g_1(|\bq|)}{ (g_1(|\bq|)^2 +g_0(|\bq|)^2)^{1/2}}  
\end{align}
is solvable. Note that the bare  mass $m_0$ appears in \eqref{eq:12}.

\section{Determination of the Dressed Electron \label{dress}}

We will find a solution of the system \eqref{eq:12}, \eqref{eq:11}
 by a fixed point argument. To this end we first integrate out the
angle on the right hand side. Setting $u:=|\bp|$ and $v:=|\bq|$ we get
\begin{align}
  \label{eq:14} 
  g_0(u) &= m_0 + \frac\alpha{2\pi} \int_0^\Lambda dv \frac vu
  Q_0\left(\frac12\left(\frac uv+ \frac vu\right)\right) \frac{g_0(v)}{
    (g_1(v)^2 + g_0(v)^2)^{1/2}}\\
  \label{eq:13} 
  g_1(u) &= u + \frac\alpha{2\pi} \int_0^\Lambda dv \frac vu
  Q_1\left(\frac12\left(\frac uv+ \frac vu\right)\right)\frac{g_1(v)}{
    (g_1(v)^2 + g_0(v)^2)^{1/2}} 
\end{align} 
where, for $z>1$, 
\begin{equation}
  \label{eq:q0} Q_0(z)= \frac12\log\frac{z+1}{z-1} 
\end{equation} 
and
\begin{equation}
  \label{eq:q1} Q_1(z)=\frac z2 \log\frac{z+1}{z-1} - 1.
\end{equation}

The case of zero bare mass is particularly easy.  We can choose
$g_0=0$, in which case $g_1$ is obtained by integration. We get
\begin{align}
  \nonumber g_1(u) = &u + \frac\alpha{2\pi} \int_0^{\Lambda} dv \frac
  vu Q_1\left(\frac12\left(\frac uv+ \frac vu\right)\right)\\
  = &u + \frac\alpha{2\pi} u\int_0^{\Lambda/u} dv \, v\, \label{eq:15}
  Q_1\left(\left(1/v+ v\right)/2\right)\\
  = &u + \frac\alpha{2\pi}u\left[ \frac23 \log|(\Lambda/u)^2-1| -
    \frac13 (\Lambda/u)^2 \right.\nonumber\\
  &\left.+ \frac {\Lambda}{6u}
    \log\left|\frac{(\Lambda/u)+1}{(\Lambda/u)-1}\right| \left(3 +
      (\Lambda/u)^2 \right)\right] \nonumber
\end{align}

The function $g_1$ behaves asymptotically for small $u$ or large
$\Lambda$ as $g_1(u)=\frac\alpha{2\pi} \frac43 u\log(\Lambda/u)$.
Unfortunately, because of the $\log u$, the operator $A$ can never
look like the renormalized Dirac operator in \eqref{newdirac}. This
asymptotic expansion can be seen either from \eqref{eq:15} or else by
noting that the integrand is a continuous function, except for $v=1$,
that decays at infinity as $4/(3v)$. This follows from the large $v$
expansion
\begin{equation}
  vQ_1\left(\left(1/v+ v\right)/2\right) = \frac4{3}v^{-1}
  +\frac8{15} v^{-3 } +O(v^{-5}).  
\end{equation}

For positive bare masses we solve the system \eqref{eq:14} and
\eqref{eq:13} by a fixed point argument. To this end we define the
following set of pairs of functions, for $\epsilon, \delta >0$.
\begin{equation}
  \label{eq:17} 
  S_{\epsilon,\delta}:= \{\bg=(g_0,g_1)\, |\,
  \forall_{u\in[0,\Lambda]}\bg(u) \in[ m_0,(1+\delta)m_0]\times
  [ u,(1+\epsilon)u] \}.  
\end{equation}

Note that with the metric generated by the sup norm
$S_{\epsilon,\delta}$ is a complete metric space. Next define
$T:S_{\epsilon,\delta}\rightarrow S_{\epsilon,\delta}$ by the right
hand side of \eqref{eq:13} and \eqref{eq:14}, i.e., 
\begin{align}
   \label{eq:19} 
   T_0(\bg)(u) &= m_0 + \frac\alpha{2\pi} \int_0^\Lambda
   dv \frac vu Q_0\left(\frac12\left(\frac uv+ \frac vu\right)\right)
  \frac {g_0(v) }{(g_1(v)^2 + g_0(v)^2)^{1/2}}\\ 
   \label{eq:18}
   T_1(\bg)(u) &= u + \frac\alpha{2\pi} \int_0^\Lambda dv \frac vu
   Q_1\left(\frac12\left(\frac uv+ \frac vu\right)\right) \frac{g_1(v)
     }{ (g_1(v)^2 + g_0(v)^2)^{1/2}}.
 \end{align}

\begin{lemma}
  \label{lem} If $Y$, defined by 
  \begin{equation} \label{YYY}
    Y:=\alpha\arsinh(\Lambda/m_0)/\pi, 
  \end{equation}
  satisfies $Y<9/50$, if \ $\epsilon \geq 50Y/(9-50Y)$, and if \ 
  $\delta \geq Y/(1-Y) $, 
  then T maps $S_{\epsilon,\delta}$ into $S_{\epsilon,\delta}$.
\end{lemma}
Note that $\arsinh(x)=\log(x+\sqrt{x^2+1})$, i.e., $Y$ grows
logarithmically in $\Lambda/m_0$.
\begin{proof} 
  Obviously, $T_0(\bg)(u)\geq m_0$ and $T_1(\bg)(u)\geq u$.
  
  To bound $g_0$ we return to \eqref{eq:12} and note the bound (which
  holds in $S_{\epsilon,\delta}$)
  \begin{equation}\label{g0upper}
    \begin{split}
    g_0(|\bp|) &\leq m_0 + \frac\alpha{4\pi^2} \int_{|\bq|<\Lambda} d\bq
    \frac{1}{|\bp-\bq|^2} \frac{(1+ \delta)m_0}{ (|\bq|^2 + m_0^2)^{1/2}} \\
    &\leq  m_0 + \frac\alpha{4\pi^2} \int_{|\bq|<\Lambda} d\bq
    \frac{1}{\bq^2} \frac{(1+ \delta)m_0}{ (|\bq|^2 + m_0^2)^{1/2}} \\
    &=m_0[1+\frac\alpha\pi (1+\delta)\arsinh(\Lambda/m_0)] .
    \end{split}
  \end{equation}
  The second inequality holds because the convolution of two symmetric
  decreasing functions is symmetric decreasing.
  
  Next we turn to $g_1$, which is a bit more complicated. We split the
  integration region in \eqref{eq:11} into $A=\{\bq \, |\,
  |\bq|<2|\bp|\ \mathrm{and} \ |\bq|< \Lambda \}$ and $B=\{\bq \, |\,
  |\bq|\geq 2|\bp|\ \mathrm{and} \ |\bq|< \Lambda \}$. In region $A$
  we use $\omega_\bp\cdot\omega_\bq \leq 1$ and hence the contribution
  to $g_1$ from this region is bounded above by
  \begin{equation}
    \frac\alpha{4\pi^2} \int_{|\bq|<\Lambda} d\bq
    \frac{1}{|\bp-\bq|^2}
  \frac{(1+ \epsilon)2|\bp|}{ (|\bq|^2 + m_0^2)^{1/2}}
    \leq 2|\bp|\frac\alpha\pi (1+\epsilon)\arsinh(\Lambda/m_0)],
  \end{equation}
  for the same reason as in \eqref{g0upper}.

  In region $B$ we use that $(|\bp|^2-|\bq|^2)^2 \geq 9|\bq|^4/16$.
  We also note, for the integration over $B$, that we can take the
  mean of the integrand for $\bq$ and $-\bq$. In other words, we can
  bound this term as follows:

  \begin{equation}\label{hilf2}
    \begin{split} 
      &\frac\alpha{4\pi^2} \left|\int_{B} d\bq
        \frac{\omega_\bp\cdot\omega_\bq}{|\bp-\bq|^2}
        \frac{g_1(|\bq|)}{ (g_1(|\bq|)^2 + g_0(|\bq|)^2)^{1/2}}\right|\\
      \leq &\frac\alpha{8\pi^2} \left|\int_{B} d\bq
        \left(\frac{\omega_\bp\cdot\omega_\bq}{|\bp-\bq|^2}
          -\frac{\omega_\bp\cdot\omega_\bq}{|\bp+\bq|^2}
        \right)\frac{g_1(|\bq|)}{ (g_1(|\bq|)^2 + g_0(|\bq|)^2)^{1/2}}\right|\\
      \leq &\frac\alpha{8\pi^2} \int_{B} d\bq
      \frac{4|\bp||\bq|}{(|\bp|^2-|\bq|^2)^2}\frac{(1+\epsilon)|\bq|
        }{
        (|\bq|^2+m_0^2)^{1/2}}\\
      \leq &\frac{8\alpha(1+\epsilon)}{9\pi^2}|\bp| \int_{B} d\bq
      \frac{1}{|\bq|^2}\frac{ 1}{ (|\bq|^2+m_0^2)^{1/2}}\\
      \leq &\frac{32\alpha(1+\epsilon)}{9\pi} |\bp|
      \arsinh(\Lambda/m_0).
    \end{split} 
  \end{equation}
  Thus, we obtain the bound
  \begin{equation}\label{aaa}
    g_1(|\bp|) \leq |\bp| \{1+\frac{ 50 \alpha}{9\pi} (1+\epsilon) 
    \arsinh(\Lambda/m_0)\}.
  \end{equation} 
\end{proof}

\begin{theorem}
  \label{th:contract}
  If $Y:=\alpha\arsinh(\Lambda/m_0)/\pi< 9/50$, if \ $(2+\epsilon
  +\delta )Y<1$ and if \ $\epsilon, \delta$ satisfy the conditions of
  Lemma \ref{lem} then $T$ is a contraction.
\end{theorem} 
\begin{proof} 
  Thanks to Lemma \ref{lem} we only need to establish the contraction
  property.  We first note that for positive real numbers $x,y, \tilde
  x, \tilde y$
  \begin{equation}\label{eq:35a}
    \left|\frac{x}{(x^2+y^2)^{1/2}}
      -\frac{\tilde{x}}{(\tilde{x}^2+\tilde{y}^2)^{1/2}}\right|\leq
    \frac{|\eta|}{\xi^2+\eta^2}((x-\tilde{x})^2+(y-\tilde{y})^2)^{1/2}
  \end{equation} 
  where $(\xi,\eta)$ is some point on the line between $(x,y)$ and
  $(\tilde{x},\tilde{y})$.
  
  Thus we get 
  \begin{equation}\label{eq:38} 
    \begin{split}
      &|[T_0(g)-T_0(\tilde{g})](u)|+|[T_1(g)-T_1(\tilde{g})](u)|\\
      \leq&\frac{\alpha}{2\pi}\int^\Lambda_0dv\frac{v}{u}
      \left[Q_0\left(\frac{1}{2}\left(\frac{u}{v}+\frac{v}{u}\right)\right)
        \left|\frac{g_0(v)}{\sqrt{g_0(v)^2+g_1(v)^2}}
          -\frac{\tilde{g}_0(v)}{\sqrt{\tilde{g}_0(v)^2+\tilde{g}_1(v)^2}}
        \right|\right.\\
      &\left.\phantom{\frac{\alpha}{2\pi}\int^\Lambda_0dv}
        +Q_1\left(\frac{1}{2}\left(\frac{u}{v}+\frac{v}{u}\right)\right)
        \left|\frac{g_1(v)}{\sqrt{g_0(v)^2+g_1(v)^2}}
          -\frac{\tilde{g}_1(v)}{\sqrt{\tilde{g}_0(v)^2
              +\tilde{g}_1(v)^2}}\right|
      \right] \\
      \leq &\frac{\alpha}{2\pi}
      \int^\Lambda_0dv\frac{v}{u}\left[Q_0\left(\frac{1}{2}\left(\frac{u}{v}
            +\frac{v}{u}\right)\right) \frac{(1+\epsilon)
          v}{v^2+m_0^2} +Q_1 \left(\frac{1}{2}\left(\frac{u}{v}
            +\frac{v}{u}
          \right)\right)  \frac{(1+\delta) m_0}{v^2+m_0^2}\right]\\
      &\phantom{\frac{\alpha}{2\pi}\int^\Lambda_0dv\frac vu[} \times
      \left[(g_0(v)-\tilde{g}_0(v))^2
        +(g_1(v)-\tilde{g}_1(v))^2\right]^{1/2}\\
      \leq &\frac{\alpha}{2\pi}
      \int^\Lambda_0dv\frac{v}{u}\left[Q_0\left(\frac{1}{2}\left(\frac{u}{v}
            +\frac{v}{u}\right)\right) 
        \frac{(1+\epsilon) v}{v^2+m_0^2}\right.\\
      &\ \phantom{\frac{\alpha}{2\pi} \int^\Lambda_0dv[}\left.+Q_1
        \left(\frac{1}{2}\left(\frac{u}{v} +\frac{v}{u}\right)\right)
        \frac{(1+\delta) m_0}{v^2+m_0^2}\right]
      \|\mathbf{g}-\tilde{\mathbf{g}}\|\\
      \leq &\frac{\alpha}{2\pi}
      \int^\Lambda_0dv\frac{v}{u}Q_0\left(\frac{1}{2}\left(\frac{u}{v}
            +\frac{v}{u}\right)\right) 
        \frac{(1+\epsilon)v+(1+\delta) m_0}{v^2+m_0^2}
      \|\mathbf{g}-\tilde{\mathbf{g}}\|\\
      \leq&(2+\epsilon + \delta)Y\|\mathbf{g}-\tilde{\mathbf{g}}\|
    \end{split}
  \end{equation}  
  where $Y=\alpha\arsinh(\Lambda/m_0)/\pi$ as in Lemma \ref{lem}.  In
  the last line we have simply noted that the integrals are smaller
  than the corresponding integral in \eqref{g0upper}.  
\end{proof}
\begin{corollary} 
  \label{cor} The map $T$ has a unique fixed point, if  $Y< 1/7$. 
\end{corollary} 
\begin{proof}
  Take $\epsilon=50Y/(9-50Y)$ and $\delta=Y/(1-Y)$ so that
  $T$ maps $S_{\epsilon,\delta}$ into itself. Then, if $Y<Y_0$, the
  contraction condition will be satisfied.  \end{proof}

For $\alpha=1/137$, i.e., the physical value of the fine structure
constant -- the condition $Y< 1/7$ is fulfilled for $\Lambda /m\leq
e^{18}$.

\section{Properties of the Renormalized Hamiltonian\label{s:4}}
 
What we have shown up to now is that the bare Hamiltonian in
\eqref{eq:1} is equivalent, apart from some additive constants, to a
renormalized Hamiltonian. This renormalized Hamiltonian has the form
\eqref{ordered}, but the quadratic term is only approximately the one
given in \eqref{newdirac}. What takes the place of $D_{Z,m}$ is given
in \eqref{eq:8} and \eqref{eq:10}. We see immediately from
\eqref{eq:14}, \eqref{eq:13} that the solution is a continuous pair of
functions. We also see that as long as $m_0$ is not zero the functions
$g_0$ and $g_1$ are positive and behave properly for small $|\bp|$,
i.e., $g_0$ is constant and $g_1$ is proportional to $|\bp|$.  To
relate this to $D_{Z,m}$ we first factor out the constant
$\lim_{u\rightarrow0+}(g_1(u)/u)$ and call this $Z^{-1}$. As is
evident from \eqref{eq:11}, this constant $1/Z$ is larger than one ---
as it should be.

The next thing to verify is that $Zg_0(0)$ is bigger than $m_0$, since
this is the renormalized mass $m$ that appears in \eqref{newdirac}. In
other words, we have to verify that
\begin{equation} \label{eq:mass}
\frac{m}{m_0}Z^{-1}=\frac{g_0(0)}{m_0}> \lim_{\bp\rightarrow 0}\frac
  {g_1(|\bp|)}{|\bp|} = Z^{-1}.  
\end{equation} 
We shall, in fact, prove more than this:
\begin{theorem}
  \label{m01} Assuming that $Y<1/7$, the unique solution to the
  equation \eqref{eq:12} and \eqref{eq:11} mentioned has the property
\begin{equation}
    \label{eq:m01} \frac{g_0(|\bp|)}{m_0}> \frac {g_1(|\bp|)}{|\bp|}
  \end{equation} 
  for all $\bp\neq \boldsymbol{0}$.
\end{theorem}
\begin{proof}
  Define $\tilde S_{\delta,\epsilon}$ to be the subset of
  $S_{\delta,\epsilon}$ on which \begin{equation}
    \label{hilf} \frac{g_0(|\bp|)}{m_0}\geq \frac {g_1(|\bp|)}{|\bp|}
  \end{equation} for all $\bp\neq \boldsymbol{0}$. If we can show that
  $T$ also leaves $\tilde S_{\delta,\epsilon}$ invariant, then we have
  shown the wanted inequality \eqref{eq:m01} on the solution in
  $S_{\delta,\epsilon}$ (by the uniqueness of the solution on
  $S_{\delta,\epsilon}$) except for the possibility that equality also
  can occur. This is so, since we can apply the fixed point argument
  not only to $S_{\delta,\epsilon}$ but also to $\tilde
  S_{\delta,\epsilon}$.

  Showing that \eqref{hilf} holds is -- according to \eqref{eq:14} and
  \eqref{eq:13} -- equivalent to showing that 
  \begin{multline}
    \label{14} \int_0^\Lambda dv \frac vu Q_0\left(\frac12\left(\frac
        uv+ \frac vu\right)\right) \frac{g_0(v)/m_0 }{ (g_1(v)^2 +
      g_0(v)^2)^{1/2}} \\> \int_0^\Lambda dv \frac vu
    Q_1\left(\frac12\left(\frac uv+ \frac vu\right)\right)\frac{g_1(v)/u
      }{ (g_1(v)^2 + g_0(v)^2)^{1/2}}
  \end{multline} 
  holds. Now, using the Inequality \eqref{hilf} and the fact that the
  factors with the roots are monotone functions $g_0$ and $g_1$
  respectively, it is enough to show that
  \begin{multline}\label{140}
    \int_0^\Lambda dv \frac vu Q_0\left(\frac12\left(\frac uv+ \frac
        vu\right)\right) \frac{1 }{ (m_0^2 + v^2)^{1/2}}\\
    > \int_0^\Lambda dv \frac vu Q_1\left(\frac12\left(\frac uv+ \frac
        vu\right)\right)\frac{v/u }{ (m_0^2 + v^2)^{1/2}}
  \end{multline}
  holds. To proceed, we now compare the integrands pointwise. Using
  the explicit expressions \eqref{eq:q0} and \eqref{eq:q1} for $Q_0$
  and $Q_1$, and with $t=v/u$, we will have shown that the integrand
  of the left hand side pointwise majorizes the one on the right hand
  side of \eqref{140} if 
  \begin{equation}
    \label{141} \log\left|\frac{t+1}{ t-1}\right| < \frac{2t}{ |t^2-1|}
  \end{equation} 
  holds for all $t\neq 1$. By symmetry, we only have to consider
  \eqref{141} for $t>1$.  To this end we exponentiate \eqref{141}
  $$\frac{t+1}{ t-1} < \exp\left(\frac{2t}{ t^2-1}\right)$$ and expand the
  exponential up to second order. (Note that this gives a lower bound
  on the exponential because the argument of the exponential function
  is positive.) I.e., it suffices to show that 
  \begin{equation}
    \label{142} \frac{t+1}{ t-1} < 1 + \frac{2t}{ t^2-1} + 2 
   \frac{t^2}{(t^2-1)^2} 
  \end{equation} 
  which follows by direct computation.

  Having established \eqref{m01} we see that \eqref{142} indeed gives
  strict inequality in \eqref{m01} for the unique fixed point in
  $\tilde S_{\delta,\epsilon}$.  
\end{proof}

\section{Interpretations of our Results}
What we have done is start with the `bare' Hamiltonian
$\mathbb{H}^{\rm bare}$ in \eqref{eq:1}, in which the interaction is
the closest analog to the classical electrostatic energy of a field.
After some analysis, we found in section 3 that $Z$ and $m$ could be
uniquely chosen so that for momenta much less than the ultraviolet
cutoff $\Lambda$, $\mathbb{H}^{\rm bare} = \mathbb{H}^{\rm
  ren}_\alpha$ plus a well defined infinite constant. Of course the
field $\Psi$ in the two cases is different. {\it Formally}, they
differ by a Bogolubov transformation, but in fact an inequivalent,
$\alpha$ {\it dependent}, representation of the CAR is needed.  In
section 4 we showed that not only is $Z^{-1} m > m_0$ but more is
true, namely $m>m_0$, and this is comforting physically.

We emphasize that we have not `integrated out' any field variables.
Our new Hamiltonian $\mathbb{H}^{\rm ren}_\alpha$ is the same as the
original one --- on a purely formal level. It is suggestive,
nevertheless, that a good deal of the electrostatic energy has now
been incorporated into the leading term $\int d^3x
:\Psi^*(x)D_{Z,m}\Psi(x):$ and that the remaining interaction is
somehow less important than the original one. It is, after all, normal
ordered, which means it vanishes on one-electron states, if we define
such states by $\Psi^* |0 \rangle$, where $|0\rangle$ is the vacuum of
the {\it new} $\Psi$.  While this makes sense perturbatively, it is,
however, misleading because the new vacuum (the state of lowest
energy) is surely not the obvious choice $|0\rangle$.

If we drop the new interaction term $\mathbb{W}^\mathrm{ren}_\alpha$
we are left with $\int d^3x :\Psi^*(x)D_{Z,m}\Psi(x):$ as our
Hamiltonian. Unfortunately this is not the Hamiltonian of a Dirac
operator (even at low momenta) because of the factor $Z^{-1}$. We have
called this a wave function renormalization, but that is not really in
the spirit of renormalizing the one-electron states (which is what is
usually done) and is, instead, a renormalization of an operator.  One
school, (K\"all\'en \cite{Kallen1958}) thinks it is proper to speak of
a renormalized $\Psi$ by incorporating a factor $Z^{-1/2}$ into $\Psi$,
but this changes the anti-commutation relations! Note that this
formulation requires renormalizing  the bare mass $m_0$ to 
$g_0(\boldsymbol{0})/ Z$ and renormalizing the bare charge $e$ to $e/Z$.

Another point of view is to regard $Z^{-1}m$ as the physical mass and
to change $Z^{-1}\bp$ into $\bp$ by a unitary transformation (which is
nothing other than a change of length scale).  This has the
disadvantage of changing the speed of light or Planck's constant. It
also would mean a different scale change for particles of different
mass, e.g., muons. As opposed to K\"all\'en's procedure, this requires
renormalizing the mass from $m_0$ to $ g_0(\boldsymbol{0})$ only, but
there is no need for charge renormalization.

Another possibility is to bring out the factor $Z^{-1}$ as a
multiplier of the whole Hamiltonian, which would mean changing the
fine structure constant to $\alpha Z$, i.e., a \textit{charge}
renormalization but now from $e$ to $e/Z^{1/2}$ only. The obvious
problem here is that since the Hamiltonian is the generator of time
translation, this means a change of the time scale (which, again,
depends on the particle in question).

Doubtless, different people will have different opinions about these
matters. We do not wish to commit ourselves to any point of view.  But
it is our opinion that the construction of
$\mathbb{H}^\mathrm{ren}_\alpha$ is a significant piece of the puzzle
of constructing a nonperturbative QED.

\textsl{Acknowledgement}: The authors thank Dirk Hundertmark for a
useful discussion. Financial support of the European Union, TMR grant
FMRX-CT 96-0001, the U.S. National Science Foundation, grant
PHY98-20650, and NATO, grant CRG96011 is acknowledged.

\end{document}